\documentclass{ws-jai}
\usepackage{longtable}
\usepackage{subfigure}

\begin{document}

\catchline{}{}{}{}{} 

\markboth{Jack Hickish et al.}{A Decade of Developing Radio-Astronomy
Instrumentation using CASPER Open-Source Technology}

\title{A Decade of Developing Radio-Astronomy Instrumentation\\using CASPER Open-Source Technology}

\author{Jack~Hickish$^{1,\dagger}$,
Zuhra~Abdurashidova$^1$,
Zaki~Ali$^2$,
Kaushal~D.~Buch$^3$,
Sandeep~C.~Chaudhari$^3$,
Hong~Chen$^2$,
Matthew~Dexter$^1$,
Rachel~Simone~Domagalski$^{1,4}$,
John~Ford$^{5,6}$,
Griffin~Foster$^{7,8}$,
David~George$^7$,
Joe~Greenberg$^9$,
Lincoln~Greenhill$^{10}$,
Adam~Isaacson$^7$,
Homin~Jiang$^{11}$,
Glenn~Jones$^{12}$,
Francois~Kapp$^7$,
Henno~Kriel$^7$,
Rich~Lacasse$^9$,
Andrew~Lutomirski,
David~MacMahon$^1$,
Jason~Manley$^7$,
Andrew~Martens$^7$,
Randy~McCullough$^5$,
Mekhala~V.~Muley$^3$,
Wesley~New$^7$,
Aaron~Parsons$^2$,
Daniel~C.~Price$^2$
Rurik~A.~Primiani$^{10}$,
Jason~Ray$^5$,
Andrew~Siemion$^{2,13,14}$,
Verees\'e~Van~Tonder$^5$,
Laura~Vertatschitsch$^{15}$,
Mark~Wagner$^{2,16}$,
Jonathan~Weintroub$^{10}$,
Dan~Werthimer$^2$,
on~behalf~of~the~CASPER~collaboration
}

\address{
$^1$Radio Astronomy Laboratory, UC Berkeley, Berkeley, CA 94720, USA \\
$^2$Department of Astronomy, UC Berkeley, Berkeley, CA 94720, USA \\
$^3$Digital Backend Group, Giant Metrewave Radio Telescope, NCRA-TIFR, Pune, 410504, India \\
$^4$Dunlap Institute for Astronomy \& Astrophysics, University of Toronto, 50 St George St, Toronto, ON, M5S 3H4, Canada\\
$^5$National Radio Astronomy Observatory, 166 Observatory Rd, Green Bank, WV 24944, USA \\
$^6$Steward Observatory, University of Arizona, Tucson, AZ 85721, USA \\
$^7$SKA Africa, 3rd Floor, The Park, Park Road, Pinelands, Cape Town, 7405, South Africa \\
$^8$Department of Physics and Electronics, Rhodes University, P.O. Box 94, Grahamstown 6140, South Africa\\
$^9$National Radio Astronomy Observatory, Central Development Laboratory,  Charlottesville, VA 22903 \\
$^{10}$Harvard-Smithsonian Center for Astrophysics, 60 Garden Street, Cambridge, MA 02138, USA \\
$^{11}$Academia Sinica, Institute of Astronomy and Astrophysics, Taiwan \\
$^{12}$Department of Physics, Columbia University, 550 W. 120th St. New York, NY 10027, USA \\
$^{13}$ASTRON, Netherlands Institute for Radio Astronomy, 7991 PD Dwingeloo, Netherlands \\
$^{14}$Radboud University, 6525 HP Nijmegen, Netherlands \\
$^{15}$Systems \& Technology Research, 600 West Cummings Park, Suite 6500, Woburn, MA 01801, USA \\
$^{16}$Nokia Bell Labs, 200 South Mathilda Avenue, Sunnyvale, CA 94086, USA \\
}

\maketitle
\corres{$^\dagger$Corresponding author. Email: \url{jackh@berkeley.edu}}

\begin{history} \received{(to be inserted by publisher)}; \revised{(to be
inserted by publisher)}; \accepted{(to be inserted by publisher)};
\end{history}

\begin{abstract}

The Collaboration for Astronomy Signal Processing and Electronics Research
(CASPER) has been working for a decade to reduce the time and cost of
designing, building and deploying new digital radio-astronomy instruments.
Today, CASPER open-source technology powers over 45 scientific instruments
worldwide, and is used by scientists and engineers at dozens of academic
institutions.  In this paper we catalog the current offerings of the CASPER
collaboration, and instruments past and present built by CASPER users and developers.
We describe the ongoing state of software development, as CASPER looks to
support a broader range of programming environments and hardware and ensure compatibility
with the latest vendor tools.

\end{abstract}

\keywords{CASPER, digital signal processing, radio-astronomy, instrumentation}

\section{Introduction}\label{sec:introduction}

Since the pioneering work of \citet{Weinreb} we
have seen a growing adoption of digital processing hardware as the foundation
on which radio telescopes are built.  Today, Central Processing Units (CPUs), Graphics Processing Units (GPUs), Field-Programmable Gate Arrays (FPGAs) and Application Specific Integrated Circuits (ASICs) power
almost all of the world's radio telescopes, and our ability to do science has
become inextricably linked with our ability to perform digital computation.
With the capability of digital processing hardware scaling exponentially with
Moore's law, the ability to leverage current technology by reducing the
design-time of new instruments is critical in effective deployment of new
radio-astronomy instruments.

The Collaboration for Astronomy Signal Processing and Electronics Research
(CASPER\footnote{\url{https://casper.berkeley.edu}}) puts
\emph{time-to-science}, the time between conception of an instrument and its
deployment, as a central figure of merit in instrument design. CASPER works to
minimize time-to-science by developing and supporting open-source,
general-purpose hardware, software libraries and programming tools which allow
rapid instrument design, straightforward upgrade cycles and reduced engineering time and cost.

CASPER hardware and software now powers over 45 radio-astronomy
instruments worldwide (see Tables~\ref{table:casper-instruments-spectrometers},
\ref{table:casper-instruments-mkids},
\ref{table:casper-instruments-correlators}) including some of the largest, most
advanced telescopes ever built, such as the upcoming MeerKat
Array\footnote{\url{http://public.ska.ac.za/meerkat/meerkat-schedule}}, the
newly commissioned Five-hundred metre Spherical Aperture Telescope (FAST,
\citet{fast}), and the Robert C. Byrd Green Bank Telescope \citep{vegas}. This paper
provides an update on the state of the collaboration, which is in the process of releasing
two new FPGA boards to the radio-astronomy community and has recently overhauled
its FPGA-programming toolflow to improve future extensibility and support the latest Xilinx software.

We first summarize the design philosophy of CASPER in
Section~\ref{sec:CASPER-philosophy}.  In Section~\ref{sec:Hardware} we describe
currently available CASPER hardware offerings, including the range of
digitizers developed and supported by CASPER. Key to CASPER's success are the
programming infrastructure, firmware libraries and software support provided by the
collaboration, which we overview in Sections~\ref{sec:toolflow}, \ref{sec:libraries}, and \ref{sec:Software}, respectively.  In
Section~\ref{sec:Deployments} we document the extensive and wide-ranging
applications to which CASPER hardware and design-tools have been applied.
Finally, we describe the future direction of, and challenges faced by, the CASPER
collaboration in Section~\ref{sec:Future}, with concluding remarks in
Section~\ref{sec:Conclusions}.

\section{The CASPER Philosophy} \label{sec:CASPER-philosophy}

The CASPER philosophy is that minimizing time-to-science is a priority when designing instrumentation. CASPER promotes open-source hardware, software and programming tools, which can be collectively developed by the community and re-used in multiple experiments to best leverage the cost and time of development.
The CASPER philosophy advocates keeping the hardware development efforts of an instrument as low as possible, and achieving cost efficiency through regular upgrades, allowing instruments to best exploit Moore's law.

This is in contrast to the way radio telescopes have been built in the past.
Large instruments, for example the ALMA correlator \citep{alma-correlator} and
eVLA's ``WIDAR'' correlator \citep{evla}, have been constructed using
custom-designed ASICs, specialized
backplanes, and cable-based interconnect in an attempt to maximize compute
density, and minimize power and cost. While there are undoubtedly arguments to be made for specialization of systems at this scale, such projects require large
hardware development budgets and timescales, and typically result in complex
and specialized instruments which lag behind the current state-of-the-art when they are deployed and are expensive to upgrade without significant
re-investment of time and money. Moreover, this development effort often does not have as significant an impact on the astronomical field as it could have, as specialization tends to limit the utility of the designed system for other applications.

In contrast, CASPER's design philosophy allows groups with the expertise and
budgets to develop hardware to benefit the entire community. To this end, the collaboration
is responsible for the development of several generations of open-source FPGA hardware which are compatible with an array of analog-to-digital converters (ADCs) and digital-to-analog converter (DAC) modules. These boards are accompanied by a suite of open-source parameterized libraries, designed to cater for the wide-ranging needs of the radio-astronomy community, and an FPGA-programming toolflow enabling portability of designs between generations of hardware.

\subsection{Modularity}

To a large extent real-time processing tasks in a modern
radio-astronomy instrument can be broken up into common parts; namely:
\begin{enumerate}
    \item Digitization of the analog sky signal, which may occur after
    some variety of analog down-conversion process. In a multi-receiver telescope, the digitization process can be parellelized over signals from multiple feeds.
    
    \item Channelization of the digitized signal into a discrete number of
    frequency bins, which is accomplished using an FFT-based filterbank
    (see, for example, \citet{specbook}). The channelization procedure may also be parallelized over signals from different feeds.
    
    \item Combination of signals from multiple antennas either via weighted addition (i.e., beamforming) or via multiplication (i.e.,
    correlation). In either case, it is possible to easily parallelize processing
    over multiple frequency channels.
\end{enumerate}

The details associated with individual instruments vary widely depending
on telescope and application. For example, digitization bandwidth and sample precision
is strongly dependent on observing frequency and analog front-end
characteristics. Some telescopes are implemented with
multiple stage channelizers, or overlapping filterbanks. Arrays will contain a varying number and configuration of antennas and may require correlators with features such as fringe-stopping and delay-tracking.
However, the recognition that
processing can effectively be divided into modular units, processing data
from only certain antennas (in the case of digitization and channelization) and
 processing data from only certain frequency channels (as is the case for
correlation and beamforming) opens up the opportunity to utilize
general-purpose computing modules, replicated as needed to meet the computational
requirements of the instrument at hand. Where the parallelization changes from
per-antenna to per-frequency, a flexible, industry-standard interconnect
solution such as Ethernet can be used to intelligently manage the
data transpose (usually referred to as the \emph{corner turn}) between
processing modules.

This is the architecture that underpins the CASPER philosophy: simple,
general-purpose compute modules, with industry-standard, commercial
interconnect \citep{parsons-petaop, parsons2008scalable, pars05}. Such a modular architecture makes it straightforward to upgrade hardware
piecemeal as Moore's law drives increases in computational capacity of
individual modules. Flexibility of modules increases the ability of multiple
groups to share hardware, and minimizes overall cost of design and manufacture.

\begin{figure}[htb]
 \centering
 \includegraphics[width=0.8\textwidth]{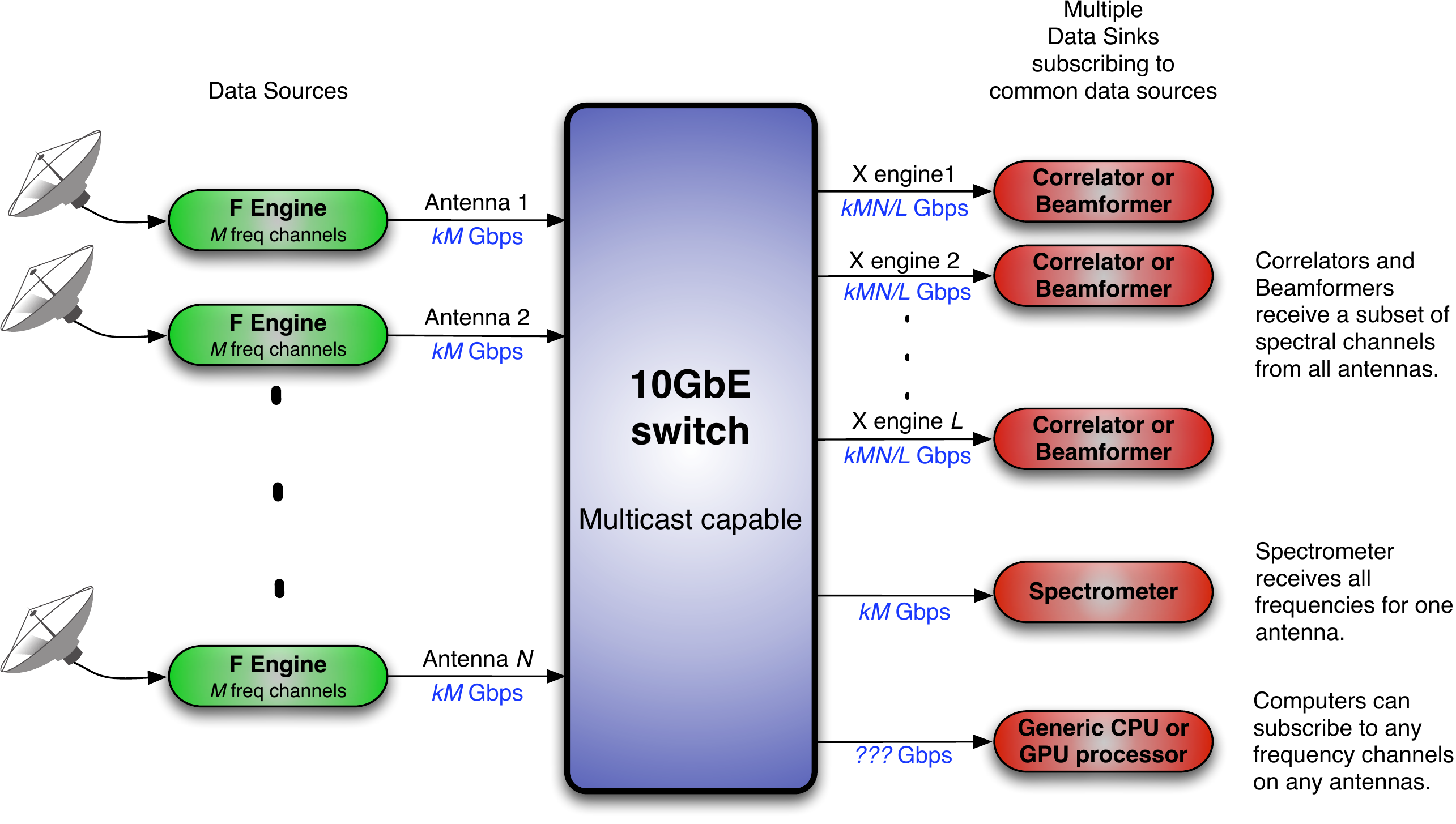}
 \caption{The canonical CASPER architecture. $N$ antennas are digitized and channelized into $M$ sub-bands using $N$ processors. These sub-bands are processed by $L$ correlation or beamforming nodes. Other users can subscribe to data streams and simultaneously implement, for example, high-resolution spectrometers or other features.}
 \label{fig:ethernet-instrument}
\end{figure}

The use of Ethernet for interconnect makes it simple for instruments to be
constructed which consist of different types of data processing hardware. For example, it is possible
to utilize FPGA technology for applications where I/O-rates are critical, such
as when interfacing with digitizers, while handing off other computationally
dense parts of the processing chain to GPU clusters. A
design can even leverage the IP Multicast protocol
which allows compute nodes to subscribe to
specific data streams and process them concurrently alongside other instruments (Figure~\ref{fig:ethernet-instrument}).
Though CASPER has championed the use of multicast switches for many years, only recently has the ability to run multiple instruments simultaneously using multicast to duplicate data streams been demonstrated in an astronomy setting \citep{man14}. 

\subsection{Flexibility}

The maximum benefits of modular hardware elements are realized when they are accompanied by similarly flexible software infrastructure. Along with hardware platforms, CASPER provides a toolflow with an interface built on MATLAB, Simulink and Xilinx System Generator (XSG) which enables a designer to easily
design and target their chosen hardware platform \citep{pars05}. The interface, described in more detail in Section~\ref{sec:Software}, 
is a graphical design environment where a designer can drag and
drop computational blocks from a provided library and connect them with wires in the desired configuration.
The toolflow and library elements are designed to be intuitive to use and provide a ``One-Click'' solution from design to
bitstream, ready to upload onto a board. The flow is designed to reduce the knowledge barrier to entry and allow students and researchers unfamiliar with FPGA technology to rapidly build their own instruments.
Generic, parameterized programming libraries are also provided, aiming to provide the user with all the building blocks needed to create a digital signal processing system.

\subsection{Community}
The CASPER collaboration currently has over 500 subscribers to its maillist, where users and developers share knowledge, instrument designs and troubleshooting advice. CASPER holds a yearly workshop usually with around 100 attendees, where users and developers may present their instrumentation work, as well as attend tutorial sessions. Groups thinking of deploying instruments based on CASPER hardware are keenly encouraged to visit experienced users who work at various academic institutions worldwide.
The collaboration places a particular emphasis on encouraging students -- including those who may lack formal training in DSP and FPGA design -- to take on roles in instrument design projects.

The collaboration also endeavors to maintain strong links with non-CASPER instrumentation groups, in order to minimize duplication of development efforts. 

\subsection{Design re-use}
The time-to-science metric of an experiment is strongly influenced by the ability to
reuse designs. This reuse extends to both an instrument's hardware, as well as its signal processing and control software infrastructure.
The collaboration provide a number of simple template spectrometer and correlator designs\footnote{See \url{https://casper.berkeley.edu/Tutorials}} which many researchers use as a starting point for their instruments.

Developers designing new hardware, software, and library modules are encouraged to make their work available to the wider community via contributions to the collaboration's central software repositories, and open-sourcing of hardware designs. Many projects choose to make the entirety of their firmware and software available on public repositories, providing a range of reference designs on which future instruments can be based. 

A case-study for successful design reuse can be found at NRAO's Green Bank Observatory, a core user and developer of CASPER infrastructure.
The Green Bank Ultimate Pulsar Processing
Instrument~\citep[GUPPI]{guppi}, which uses the CASPER DSP libraries and hardware
computer systems has been duplicated both locally in Green Bank to process signals
from the retired 140 ft telescope, and later as
the primary pulsar machine at the Arecibo Observatory.  Both of these
machines were installed with no engineering effort, and a modest
effort from the computer system administrators and pulsar scientists.
\emph{In the past, these clonings would have been much more time- and
  labor-intensive, hence more expensive.}

The next generation of designs to be created in Green Bank from CASPER
were all based on the ROACH family of configurable signal processors.
As noted in Table~\ref{table:casper-instruments-spectrometers}, several ROACH based spectrometers
are deployed in Green Bank performing various functions for PI-based
science~\citep[for example]{skynet}.  These all use common
hardware and firmware, and a great deal of the software for processing
and analyzing the output is common as well.  None of the PI-based
science would have been feasible without the ease of reuse of the
ROACH designs.

Green Bank's current facility instrument -- the VEGAS
spectrometer~\citep{chennamangalam2014gpu} -- was built with reused GUPPI
software, reused CASPER gateware libraries, and custom gateware
blocks that were subsequently added to the CASPER libraries for others
to use.  Calibration algorithms for the VEGAS high-speed ADCs, a key component of the system, were gleaned from other
CASPER users~\citep{adc5g}.
The entire VEGAS spectrometer has been cloned and enhanced for
deployment at telescopes in China as well
as in other spectrometers in Green Bank.  The enhancements from these
other deployments have since been retrofitted to the original VEGAS system.

%
%

\subsection{History}
The original CASPER toolflow was inherited from the Berkeley Wireless Research Center's BEE2 project\footnote{\url{http://bee2.eecs.berkeley.edu/wiki/BEE2wiki.html}} which created the ``bee\_xps'' flow and BORPH operating system for FPGA-based platforms \citep{borph1, borph2, borph-thesis}. This flow was based around a MATLAB object oriented framework for generating a Xilinx Embedded Development Kit (EDK) project, along with associated constraints, and compiling it into a \emph{.bof} file -- a container for the bitstream and its meta-data. Users would interface with bee\_xps via MATLAB's Simulink schematic entry and simulation tool, in which FPGA designs could be created by connecting blocks with various logical functions.

Having compiled a design to a .bof file using bee\_xps, users could directly execute code on an FPGA-platform running the BORPH operating system, as if the hardware designs were software processes. Software and hardware could then communicate via standard UNIX file pipes and a virtual file system allowing access to FPGA memory resources.

The bee\_xps flow, discussed more in Section~\ref{sec:toolflow} in the context of current developments, provided a simple and intuitive way to design DSP systems. When coupled with astronomy-centric DSP libraries (Section~\ref{sec:libraries}) bee\_xps proved to be a powerful tool for designing radio-astronomy instruments and is the foundation on which CASPER infrastructure is built.

\section{CASPER Hardware} \label{sec:Hardware}

While CASPER advocates the use of a variety of hardware platforms, over the past decade the collaboration has focused its efforts on designing FPGA-based hardware and ADC/DAC daughter-boards\footnote{See \url{https://casper.berkeley.edu/wiki/Hardware}}. Originally utilizing processing hardware developed at the Berkeley Wireless Research Center such as the Interconnect Break-out Board (iBOB) and Berkeley Emulation Engine 2 (BEE2, \citet{bee2}) the collaboration later began designing their own platforms to best meet the needs of the radio-astronomy community. An overview of the hardware specifications of the most recent five CASPER boards is given in Table~\ref{table:fpga-hardware}. A dozen different ADC and DAC add-on cards are available for these platforms (Table~\ref{table:adc-hardware}). A brief overview of available FPGA platforms, focusing on the latest-generation SKARAB and SNAP boards, is given below.

\begin{table}[htb]
\caption{A decade of CASPER FPGA hardware.}
\label{table:fpga-hardware}
\centering
\begin{tabular}{lccccc}
 & iBoB & ROACH & ROACH2 & SNAP & SKARAB \\
\hline
Year Available    & 2005     & 2009     & 2010        & 2016        & 2016               \\
Logic cells       & 53K      & 94K      & 476K        & 162-406K    & 693K               \\
DSP slices        & 232      & 640      & 2016        & 600-1540    & 3600               \\
BRAM capacity     & 4.2 Mb   & 8.8 Mb   & 38 Mb       & 11-28 Mb    & 53 Mb              \\
SRAM capacity     & 2x18 Mb  & 2x36 Mb  & 4x144 Mb    & -           & -                  \\
SRAM bandwidth    & 9 Gb/s   & 43 Gb/s  & 200 Gb/s    & -           & -                  \\
DDR capacity (max)& -        & 1x8 Gb   & 1x16 Gb     & -           & -                  \\
DDR bandwidth     & -        & 38 Gb/s  & 50 Gb/s     & -           & -                  \\
HMC capacity      & -        & -        &             & -           & $<$8x32 Gb         \\
HMC bandwidth     & -        & -        &             & -           & $<$8x30 Gb/s       \\
Ethernet ports    & 2x10 GbE & 4x10 GbE & 8x10 GbE    & 2x10 GbE    & $<$16x40 GbE       \\
ADC/DAC support   & 2xZDOK   & 2xZDOK   & 2xZDOK      & 1xZDOK, 3xHMCAD1511 & 4xMegarray \\
\end{tabular}
\end{table}

\begin{table}[htb]
\caption{CASPER ADC and DAC boards developed for use with FPGA-platforms.}
\label{table:adc-hardware}
\centering
\begin{tabular}{lccccc}
Name               & Type    & Bits  & Analog   & Sample Rate & Interface \\
                   &         &       & Channels & (MS/s)      & \\
\hline
ADC2x1000-8 (iADC) & ADC     & 8     & 1/2    & 2000/1000    & Z-DOK \\
ADC1x3000-8        & ADC     & 8     & 1      & 3000         & Z-DOK \\
ADC64x64-12        & ADC     & 12    & 64     & 64           & dual Z-DOK \\
ADC4x250-8 (QuADC) & ADC     & 8     & 4      & 250          & Z-DOK \\
ADC2x550-12        & ADC     & 12    & 2      & 550          & Z-DOK \\
ADC2x400-14        & ADC     & 14    & 2      & 400          & Z-DOK \\
KatADC             & ADC     & 8     & 1/2    & 3000/1500    & Z-DOK \\
ADC1x5000-8 \citep{2014PASP..126..761J}       & ADC     & 8     & 1/2    & 5000/2500    & Z-DOK \\
ADC16x250-8 \citep{adc10g}                    & ADC     & 8     & 4/8/16 & 1000/500/250 & Z-DOK \\
ADC1x10000-4       & ADC     & 4     & 1      & 10000        & Z-DOK \\
DAC2x1000-16       & DAC     & 16    & 2      & 1000         & Z-DOK \\
MUSIC              & ADC/DAC & 12/16 & 2/2    & 550/1000     & dual Z-DOK \\
\end{tabular}
\end{table}

\paragraph*{iBOB}
Designed in collaboration with the Berkeley Wireless Research Center and the UC Berkeley SETI group, the iBOB is a Xilinx Virtex II platform designed to interface ADC cards with a commercial 10~Gb/s Ethernet network. Approximately 100 iBoBs have been delivered to the astronomy community \citep{private-mo} and powered instruments at the Parkes telescope \citep{2010MNRAS.409..619K}, Robert C Byrd Green Bank Telescope \citep{guppi} as well as the SETI instrument SERENDIP V.v \citep{seti}.
\paragraph*{ROACH}
The ROACH architecture built on the single FPGA architecture of the iBoB and added a control processor, enhanced memory and connectivity options \citep{Casp09}. The core of the ROACH is the Xilinx Virtex 5 FPGA. With around 280 boards delivered \citep{private-mo} the ROACH is the most prolific CASPER board to date, and is still supported by the current CASPER tools.
\paragraph*{ROACH2}
The ROACH2 is an update to the ROACH platform, featuring a Xilinx Virtex 6 FPGA with increased processing and IO capabilities. The ROACH2 maintained the control processor architecture of its predecessor, allowing users to increase the capability of their systems with little or no changes to their control and monitoring software. Approximately 180 ROACH2 boards have been delivered to researchers to date \citep{private-mo}.
\paragraph*{SKARAB}
The Square Kilometre Array Reconfigurable Application Board (SKARAB) is the latest generation of CASPER FPGA hardware. Unlike previous CASPER platforms, the SKARAB was designed by the South African company, Peralex\footnote{\url{www.peralex.com/product_SKARAB.html}}, according to the specifications of SKA-SA.

The SKARAB makes provision for four mezzanine card sites with each site providing an interface to 16 high-speed (10~Gb/s) serial transceivers \citep{cliff16}.
Two mezzanine cards currently exist for SKARAB: a QSFP+ Mezzanine Module which provides support for four 40Gb Ethernet interfaces, and a Hybrid Memory Cube (HMC) module providing additional memory capacity. ADCs compatible with the SKARAB mezzanine interface are an area of active research.

The SKARAB board does not include an on-board CPU, though provision has been made for the COM Express mezzanine site which can interface with an external processor via single lane PCIe \citep{Teag15}. Instead, control of the board has been implemented using a Microblaze soft processor core\footnote{\url{www.xilinx.com/products/design-tools/microblaze.html}}.

\begin{figure}
  \centering
  \subfigure[The SKARAB platform is a modular processing platform based around a Xilinx Virtex 7 FPGA with expandable mezzanine slots for memory and Ethernet ports.]{\includegraphics[height=2.6in]{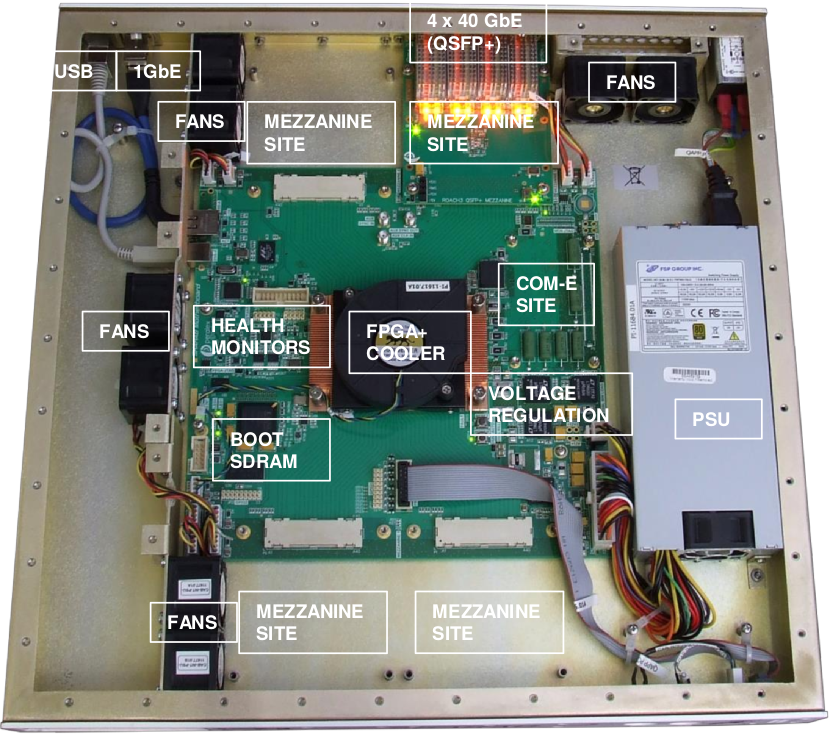}\label{fig:skarab_hw}} \hspace{1cm}
  \subfigure[The SNAP platform, developed for the HERA array, is a low-cost platform based around a Xilinx Kintex-7 FPGA. It features on-board ADCs and target applications requiring digitization, channelization and packetization of RF signals.]{\includegraphics[height=2.6in]{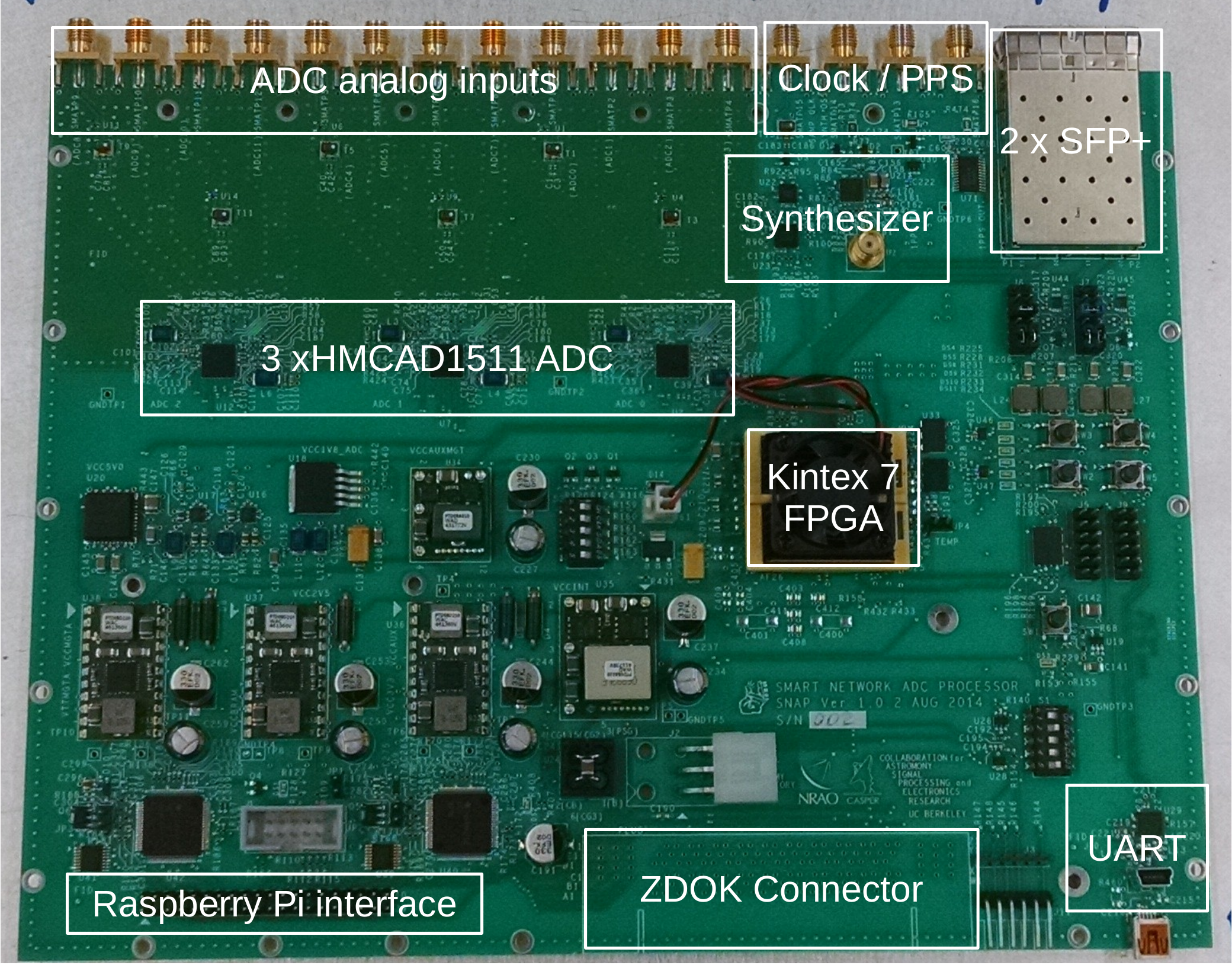}\label{fig:snap_hw}} 
   \caption{The latest two CASPER-supported boards. The SKARAB (a) and SNAP (b).} \label{fig:test}
\end{figure}

%
SKARAB boards have recently been made available to the general community, with the MeerKAT project planning to deploy 300 boards by the end of 2017. The SKARAB board is shown in Figure~\ref{fig:skarab_hw}.

\paragraph*{SNAP}
The Smart Network ADC Processor
(SNAP\footnote{\url{https://casper.berkeley.edu/wiki/SNAP}}, Figure~\ref{fig:snap_hw}) is a lightweight, next generation
FPGA platform designed primarily to perform digitization, channelization and packetization of analog signals in the Hydrogen Epoch of Reionization Array (HERA) experiment \citep{2016arXiv160607473D}. HERA requires digitization of around 700 signals at rates of 500~MS/s. In order to reduce cost and increase reliability, unlike previous CASPER platforms, the SNAP board features three on-board HMCAD1511 digitizer chips as well as an integrated synthesizer. Support for a single ZDOK interface is maintained to ensure compatibility with existing CASPER ADC daughter cards.
As with other CASPER platforms, the SNAP board is designed to be used with the 10~Gb Ethernet protocol and features two SFP+ outputs. Though lacking features of the ROACH series, such as off-chip memory and on-board CPU, the SNAP represents a flexible platform on which to implement generic RF to Ethernet digitization schemes.
The SNAP board will implement the same Microblaze-based control system as SKARAB, though users are also able to interface a CPU with the board using a simple 40-pin ribbon connector. The SNAP platform provides an interface designed to be compatible with the popular and widely available Raspberry Pi single board computer\footnote{\url{https://www.raspberrypi.org/}} which enables the SNAP to be used with software originally designed to target ROACH platforms.
%

\section{The CASPER Toolflow} \label{sec:toolflow}

Central to CASPER's wide adoption in the radio-astronomy community is the provision of a graphical toolflow which provides ``One-Click'' compile capability. Once a user has developed a design in MATLAB's Simulink environment, a single command will generate a programming file ready to be loaded onto a CASPER board. This programming file encapsulates not only the FPGA bitstream, but also meta-data about software-controllable blocks in a user's design. Coupled with the software infrastructure provided by CASPER, this allows users to load a firmware design and interact with it in real-time in an extremely straightforward and intuitive manner.
The CASPER toolflow allows users to design DSP systems without being responsible for low-level implementation details, such as ADC interfaces, or Ethernet implementations, which are handled automatically by the environment. The user is also spared the task of configuring physical design constraints such as timing requirements and FPGA pin locations, which are automatically generated based on the contents of a user's design and their target platform. Critically, the toolflow allows a complete design -- including blocks representing hardware elements such as ADC interfaces, external memory and ethernet ports -- to be trivially ported between compatible hardware platforms by changing a single top-level parameter.

The CASPER toolflow (known also as the bee\_xps flow, after it's heritage as part of the BEE project) served the CASPER community well, but had a number of drawbacks:
\begin{enumerate}
 \item Few members of the CASPER community were familiar with MATLAB Object Oriented Programming, creating an immediate barrier to users becoming toolflow developers.
 \item The bee\_xps flow is entirely reliant on MATLAB. This goes against the desires of the CASPER community for a free, open-source flow. Where aspects of the flow could in-principle be reimplemented outside of MATLAB, the monolithic nature of the bee\_xps flow made this difficult.
 \item The bee\_xps flow is strongly coupled to MATLAB's Simulink design entry tool. While this is an advantage for new users, who appreciate a graphical interface, advanced users and developers have long been requesting alternative design input methods.
 \item The bee\_xps flow uses Xilinx's ISE package, which has not been updated since 2013 and is not supported by the latest generations of Xilinx FPGAs.
\end{enumerate}

The last of these drawbacks became particularly significant with the release of the SKARAB and SNAP hardware platforms, which feature the last generation of FPGAs to support Xilinx ISE. This has led to the development of a new toolflow, nominally under the name \emph{JASPER}, designed to support the latest generations of FPGAs and alleviate some of the shortcomings of bee\_xps.
The JASPER flow has the following features:
\begin{enumerate}
 \item Written in pure Python 2.7, with minimal dependencies.
 \item Supports the same Simulink design entry method used by bee\_xps, allowing portability of existing designs to the new flow.
 \item Designed to be modular, with the design entry method de-coupled from the management of constraints and interface code which the toolflow aims to hide from the designer.
 \item The flow is also de-coupled from the FPGA vendor's compilation tools. This allows the JASPER flow to support both Xilinx ISE (required by older platforms) as well as Xilinx's new Vivado software suite required by the newest generation of FPGAs.
\end{enumerate}

The JASPER flow is currently being used for SNAP and SKARAB designs, which support Xilinx Vivado. Written in Python, this flow is designed to be easier to modify by the numerous Python-proficient developers in the CASPER collaboration.

A key change introduced in JASPER is to break up the flow into modular elements. Design entry and functional simulation is provided by a front-end (of which Simulink is currently the only supported option). Management of constraints and source code generation is conducted by JASPER's Python internals, which hand off compilation of an FPGA bitstream to a vendor-specific back-end.
 
Through this modularity JASPER aims to provide a mechanism to facilitate multiple front-end environments. This might include text-based tools utilizing high-level FPGA design packages, such as Migen or MyHDL, or graphical interfaces such as Simulink, Labview, Sci-Lab/Sci-Cos.

A separate backend allows the toolflow to utilize a variety of vendor compilers. Currently both Xilinx ISE and Vivado are supported, though in the future support for other FPGA vendors is desirable.

\section{CASPER DSP Libraries} \label{sec:libraries}



In order to quickly and easily develop new radio-astronomy instruments a number
of DSP blocks have been developed for use in CASPER board firmware. Typical
instruments such as spectrometers, beamformers, correlators, ADC recorders, and
DAC signal generators are constructed from this library of DSP blocks.

These blocks are based on the low-level logical units provided by the Xilinx
Simulink library and the generic Simulink libraries. Configurable low-level DSP
blocks are then used to build more complex, high-level DSP blocks. This
heirarchical design style enables quick and uniform logic development through
block reuse. High-level blocks include configuration parameters which are
propagated through the underlying logic. Thus,
generic blocks such as streaming FFTs and vector accumulators are included in
the library, and configured when placed into a firmware design.

Modules for the cross-correlation operation in an FX correlator have been
implemented in the library. A complex multiply and accumulate (CMAC) block for
small correlators, usually on a single board for a small number of inputs, can
be used in a matrix-style design. For large-N correlator systems distributed
across multiple boards a streaming windowed X-engine can be used \citep{parsons2008scalable, hickish14}.

A number of vector accumulator modules are in the library.
Small vectors, such as those from a spectrometer, are implemented in BRAM, while
large vectors, such as the cross-correlations of a correlator, are implemented
with QDR or DRAM memory. These accumulators include an interface making it possible to access
the results via software during runtime if desired.

Blocks for including dynamic delays in beamformer and correlator systems have
been implemented. A configurable `coarse' delay can be applied before the channelizing 
operation, while a `fine' delay can be applied post-channelization with a phase correction.
These delays can be combined with a software interface to act as a fringe-tracking module. 

Modules designed to enable debugging of instruments at run-time,
such as parameterized noise generation blocks \citep{noisegen}, and blocks
to capture snapshots of data, have also been developed.

In addition to the generic DSP blocks, a number of external interfaces known as
`yellow blocks' have been designed to interact with hardware such as
ADCs, DACs, software-accessible registers, QDR and DRAM memory, and network
interfaces. 
Within the Simulink environment a yellow block acts as a place holder for low-level interface firmware, which is included by the CASPER toolflow when a design is compiled. The toolflow is designed such that designs using these yellow blocks are easily portable between FPGA hardware generations.
Among other abstractions, these yellow blocks and their associated software control scripts handle reliable data flow across asynchronous clock boundaries, allowing users to avoid one of the more difficult aspects of FPGA design.

The main collaboration library is hosted on
github\footnote{github.com/casper-astro/mlib\_devel} with a number of
other projects creating their own forks of this library.

\section{CASPER Software} \label{sec:Software}
\begin{figure}
  \centering
  \subfigure[The Simulink environment provides blocks to interface to external hardware such as memories and ADCs. Special blocks also allow users to interact with a running FPGA design using standard CASPER software libraries.]{\includegraphics[width=0.6\textwidth]{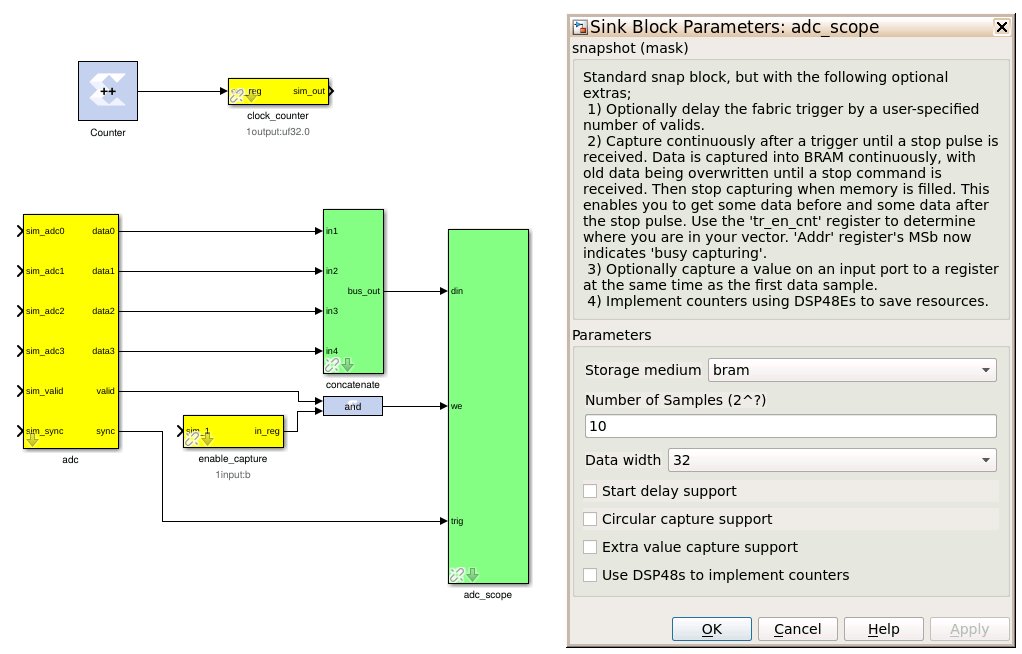}\label{fig:simulink}} \hfill
  \subfigure[Example code to interact with a running FPGA design using the \emph{corr} Python package (hosted at https://github.com/ska-sa/corr).]{ \includegraphics[width=0.35\textwidth]{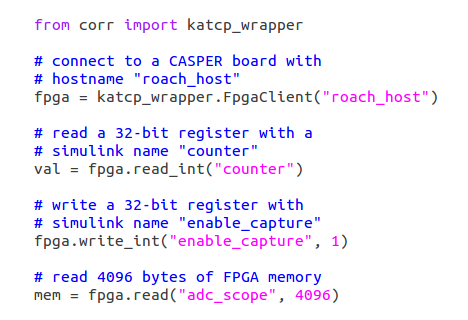}\label{fig:sw}}
  \caption{The CASPER toolflow automatically manages memory elements in an FPGA design allowing users to write intuitive software to control and monitor firmware at runtime.} \label{fig:toolflow}
\end{figure}

Key to the CASPER infrastructure is the ability to easily monitor and control
FPGA-boards via stable, extensible, software libraries. The Karoo
Array Telescope Control Protocol
(KATCP\footnote{\url{https://casper.berkeley.edu/wiki/KATCP}}), first developed for use with the ROACH platform, has accomplished this task. KATCP is a simple text-based protocol, which now has a suite of available
client libraries supporting C Language\footnote{\url{https://github.com/ska-sa/katcp_devel}}, Python\footnote{\url{https://pypi.python.org/pypi/katcp}}, Ruby\footnote{\url{http://rb-katcp.rubyforge.org/}}, and LabView\footnote{\url{ftp://ftp.cv.nrao.edu/NRAO-staff/jcastro/CASPER/LabVIEW/}}.
This CASPER software complements the toolflow environment, and allows the user to access software-accessible blocks with names and functions that can be configured using the intuitive graphical interface (Figure \ref{fig:toolflow}). 

\section{CASPER Deployments} \label{sec:Deployments}

CASPER hardware has been extensively used in a variety of astronomical instruments. Here we give a summary of known deployed systems based on CASPER FPGA, ADC, and DAC hardware. In many cases, CASPER hardware has been used to complement commodity CPU and GPU hardware processing resources - the ease with which these systems can be constructed is a key advantage of the Ethernet interconnect advocated by CASPER.

CASPER hardware has also gained some traction outside of radio-astronomy. For example, the Manastash Ridge Radar has adopted the CASPER ROACH and ROACH2 boards and toolflow to create a frequency-agile, high bandwidth passive radar system for detection of aerospace and geoscience targets \cite{vertatschitsch_2013}.  Fast sampling allows the system to digitally select illuminators of opportunity such as GPS, HDTV, or FM radio, and the high aggregate bandwidth out of the 10 GbE ports allows for simultaneous reception of these signals from a single RF front-end. Other known uses include DNA alignment searching \citep{roach-blast} and wireless neural sensor readout \citep{implants}.

\subsection{Spectrometers}

Tables \ref{table:casper-instruments-spectrometers} gives a summary of spectrometers based on CASPER hardware. Though by no means representative of all CASPER spectrometers - one example of a large-scale facility instrument is the VEGAS spectrometer \citep{vegas}.

\begin{figure}[htb]
 \centering
 \includegraphics[width=\textwidth]{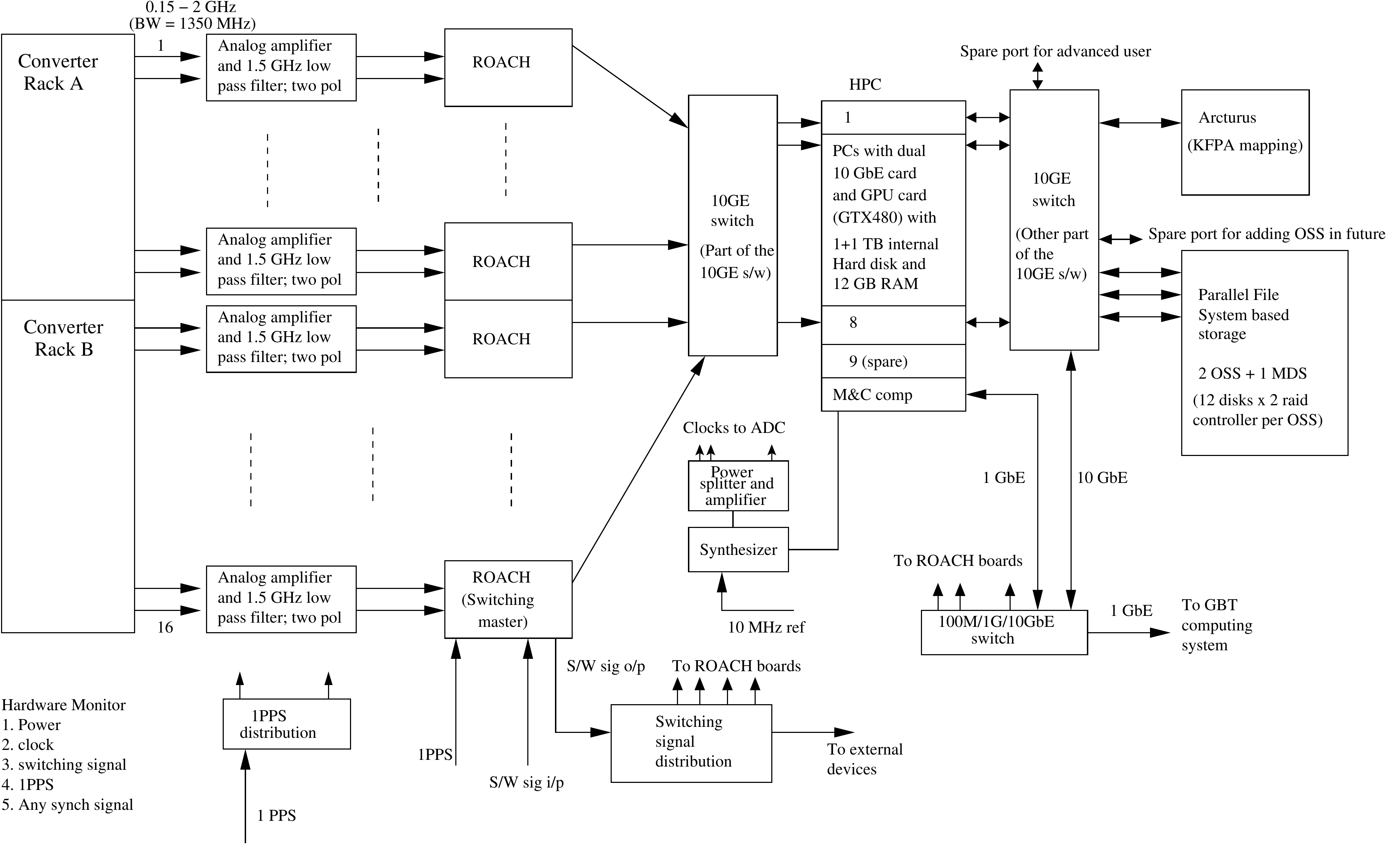}
 \caption{A block diagram of the heterogeneous VEGAS system, which digitizes signals using CASPER ADC and FPGA hardware.}
 \label{fig:vegas}
\end{figure}

The VEGAS spectrometer is based on a ROACH2 FPGA frontend and a heterogeneous computing backend comprised
of GPUs and x86-64 CPUs, as shown in
Figure~\ref{fig:vegas}. The hardware in this system provides
processing power to analyze up to 8 dual-polarization or 16
single-polarization inputs, at bandwidths of up to 1.25~GHz per
input. An aggregate of up to 10 GHz of bandwidth, dual polarization,
may be simultaneously processed with the VEGAS spectrometer.

VEGAS supports numerous processing modes\footnote{See
\url{http://www.gb.nrao.edu/vegas/modes}} which are able to generate spectra at a variety of time and frequency resolutions. High-Bandwidth modes use only the FPGA to process the data, and are used for observations over the full input bandwidth of the system with a modest frequency resolution of up to $~100$~kHz. Low-Bandwidth modes harness the power of the GPUs to create spectra with up to 20~Hz resolution
on modest---$~10$~MHz---bandwidths.  Multiple spectral windows may be placed
inside the analog bandwidth of the system, enabling high-resolution
spectroscopy of widely-spaced lines.

This same hardware is being repurposed with new firmware and control
software to provide unprecedented pulsar capability.  The new pulsar
modes will provide double the bandwidth of GUPPI, and double the
number of channels, and up to 8 dual-polarization pulsar inputs.  This
new capability, when realized, will enable the retirement of GUPPI,
now over 8 years old.

\begin{longtable}{ccp{10cm}}
  \caption{Spectrometers and packetizers powered by CASPER hardware.}\\
  Instrument & Year & Description \\
  \hline \endfirsthead
  \caption[]{continued}\\
  Instrument & Year & Description \\
  \hline \endhead
  Fly's Eye        & 2007 & High time-resolution spectrometer for the 42-dish Allen Telescope Array. Based on 11 iBOB boards with 22 ADC2x1000-8 digitizers \citep{flyseye} \\
  GUPPI            & 2009 & Pulsar processor with the Green Bank Telescope. Provides full stokes polarimetry and ethernet packetizing with up to 800 MHz BW. Based on a pair of iBOB boards with ADC2x1000-8 digitizers and a BEE2 \citep{guppi}. \\
  CASPSR           & 2009 & 400~MHz bandwith packetizer for GPU-based pulsar processing backend at the Parkes telescope. Originally implemented with an iBOB. Later upgraded to a ROACH board with ADC2x1000-8 digitizer\footnote{\url{https://astronomy.swin.edu.au/pulsar/?topic=caspsr}} \\
  BPSR             & 2009 & 400~MHz bandwidth, 13-beam fast-dump spectrometer for the Parkes multibeam receiver. Originally implemented with iBOBs. Upgraded in 2012 to use 13 ROACH boards \citep{mcmahon-thesis, 2010MNRAS.409..619K}. \\
  GAVRT            & 2009-2012 & 8~GHz instantaneous bandwidth transient capture buffer with real-time incoherent dedispersion trigger. Implemented with 8 iBOBs, 16 ADC2x1000-8s and a BEE2 \citep{jon10, JonesDSS28}. \\
  SERENDIP V.v     & 2009+ & 200~MHz spectrometer with 1.5~Hz resolution, fed from the Arecibo L-band Feed Array. SERENDIP V.v was deployed in 2009 \citep{seti} and was implemented with an iBOB and BEE2. The latest generation, SERENDIP VI, uses a single ROACH2 and ADC1x5000-8 digitizer to process $>$1~GHz bandwidth. SERENDIP VI has been deployed at both the Green Bank and Arecibo Observatories. \\
  HiTREKS          & 2010 & A 104.8~MHz bandwidth total power and kurtosis spectrometer used to detect dust storm induced lightning on Mars \citep{mars-lightning}. \\
  NUPPI            & 2011 & 512~MHz digitization and flexible (32-1024 channel) channelizer for pulsar observations. Implemented using a single ROACH board with ADC2x1000-8 digitizer \citep{2014MNRAS.443.3752L}. \\
  Skynet           & 2012 & Single board ROACH-based educational spectrometer for the Green Bank 20m telescope. Provides 500MHz BW, dual-polarized input, and 1024 channels \citep{skynet}. \\
  RATTY            & 2012 & Transient / RFI Monitor for SKA-SA site monitoring implemented on a single ROACH board \citep{Foley01082016, man14}. \\
  cycSpec          & 2012 & Real-time cyclic spectrometer, deployed at Arecibo (ROACH) and GBT (ROACH2) on consecutive generations of hardware. Implements a filterbank of 128~MHz overlapping channels used to feed GPU processors \citep{cycspec}. \\
  C-BASS           & 2013 & C-Band All-Sky Survey, utilizing antennas at the Owen's Valley Radio Observatory, CA, and South African MeerKAT site. C-BASS utilizes ROACH and iADC hardware to implement a 4.5--5.5~GHz continuous comparison radiometer and polarimeter \citep{chuckles-thesis}.\\
  HIPSR            & 2014 & 400~MHz bandwidth 8192 channel spectrometer and high time-resolution system for the Parkes multibeam receiver. Based on 13 ROACH1 boards and 13 ADC2x1000-8 digitizers\footnote{\url{http://telegraphic.github.io/hipsr/overview.html}}. \\
  KuPol            & 2014 & 6 GHz bandwidth full Stokes spectrometer, implemented with 12 ROACH boards and 24 ADC2x1000-8 digitizers \citep{2013arXiv1303.2131M}. \\
  VEGAS            & 2014 & Versatile spectrometer for the GBT, providing up to 10GHz BW for 1 dual-polarized input or 1.25GHz BW for 8x dual-polarized inputs. Features wideband modes and narrowband modes with 8 digitally tuned sub-bands within the 1.25GHz BW. Implemented using 8 ROACH2 boards with ADC1x5000-8 digitizers \citep{chennamangalam2014gpu}. \\
  ALMA Phasing Project& 2014 & 8 ROACH2 system for time-tagging, ethernet packetization and VDIF (VLBI) formatting \citep{2012evn..confE..53A}. \\
  Leuschner        & 2015 & Dual-polarization, 12 MHz, 8192 channel spectrometer for UC Berkeley's Leuschner Radio Observatory. Based on ROACH1 and ADC2x1000-8\footnote{\url{https://github.com/domagalski/leuschner-spectrometer}} \\
  R2DBE            & 2015 & 4~GHz bandwidth 2-bit data recorder for VLBI with the Event Horizon Telescope. Built using a ROACH2 with a pair of ADC1x5000-8 digitizers and high-speed Ethernet output \citep{r2dbe}. \\
  DSN Transient Observatory & 2016 & Versatile signal processor for commensal astronomy during DSN data downlinks, featuring Kurtosis Spectrometer and pulse detection. Based on a pair of ROACH boards with KATADCs \citep{dsn-transient-observatory} \\
  VGOS             & 2016 & The VLBI Global Observing System utilizes 4 ROACH boards each with ADC2x1000-8 digitizers to implement a dual-polarization VLBI recording system. Ten such VGOS stations are being deployed worldwide \citep{2012ivs..conf....8H}.\\
  AVN-Ghana        & 2016 & Single dish observing mode for the African VLBI Network's Ghana antenna. A ROACH and katADC digitizer are used to implement a spectrometer with wideband (400~MHz, 0.39~MHz resolution) and narrowband (1.56~MHz, 381~Hz resolution) modes \citep{avn}.\\
  COMAP            & Development & 8~GHz, 19-beam spectrometer, with digital sideband separation. Implemented using 38 ROACH2 boards\footnote{\url{http://www.astro.caltech.edu/CRAL/projects.html}}. \\
  \label{table:casper-instruments-spectrometers}
\end{longtable}

\subsection{Kinetic Inductance Detectors}

Groups deploying Microwave Kinetic Inductance Detector (MKID) systems, which use multiple microresonators as mm/submm photon detectors, have been a key user of CASPER technologies. MKID systems are read out by generating combs of microwave tones and using them to monitor the resonant frequencies and dissipations of the resonating detectors.

\begin{figure}[htb]
 \centering
 \includegraphics[angle=90,origin=c, width=0.5\textwidth]{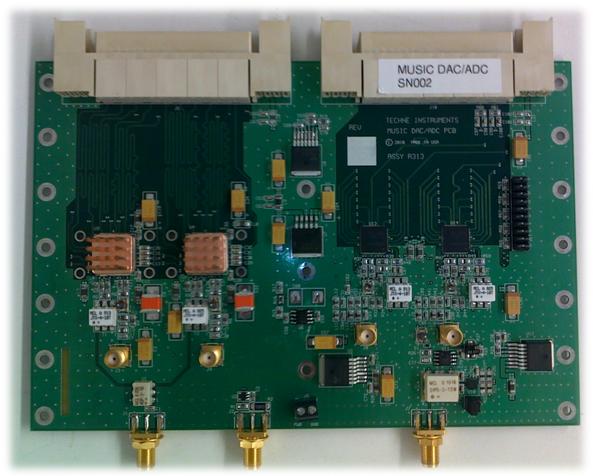}
 \caption{The MUSIC DAC/ADC board, designed to interface with ROACH hardware to provide MKID readout capability.}
 \label{fig:music-adc}
\end{figure}

Unlike other use-cases, MKID readout systems require DACs to generate output signals. A healthy collection of DAC boards is now an integral part of the CASPER ecosystem (Table~\ref{table:adc-hardware}). Various projects have shared ADC and DAC technologies, such as the MUSIC ADC/DAC card (Figure~\ref{fig:music-adc}), and taken advantage of the ability to upgrade instruments between FPGA hardware generations. A list of MKID readout systems using CASPER technologies is given in Table~\ref{table:casper-instruments-mkids}.

\begin{longtable}{ccp{10cm}}
  \caption{MKID readout systems powered by CASPER hardware.}\\
  Instrument & Year & Description \\
  \hline \endfirsthead
  \caption[]{continued}\\
  Instrument & Year & Description \\
  \hline \endhead
  Columbia MKID    & 2012 & ROACH (later ROACH2) based MKID readout system with CASPER-based tone generation, digitization and coarse channelization. Feeds non-CASPER HPC processors \citep{mccarrick_2014}. \\
  Mustang2         & 2015 & 90GHz FPA TES bolometer array, with Microwave-multiplexed SQUIDs. Based on 4 ROACH boards \citep{2016JLTP..184..460S, 2014JLTP..176..808D}.  \\
  DARKNESS         & 2016 & 10,000 MKID pixel readout system with 2~GHz bandwidth between 4-8~GHz. Implemented with 10 ROACH2 boards and custom ADC/DAC/IF cards \citep{meeker2015design}. \\
  MEC              & Development & Expansion of DARKNESS system to accommodate 20,000 pixel readout, using 20 ROACH2 boards with custom ADC/DAC/IF cards \citep{meeker2015design}. \\
  BLAST-TNG        & Development &  2.5~m Balloon-Borne Submillimeter Polarimeter with CASPER MKID readout system. Based on 5 ROACH2 boards with MUSIC-DAC/ADC cards \citep{galitzki2014balloon}. \\
  HOLMES           & Development & Electron Neutrino Mass measurement experiment with CASPER-based microwave SQUID readout system, based on 35 ROACH2 boards with MUSIC-ADC/DAC cards \citep{Alpert2015, Ferri2016179}. \\
  \label{table:casper-instruments-mkids}
\end{longtable}

\subsection{Correlators \& Beamformers}

Numerous correlators and beamformers have been built on CASPER technologies (Table~\ref{table:casper-instruments-correlators}). In particular, CASPER developers have worked to provide and maintain firmware modules for all aspects of FX correlation systems \citep{parsons2008scalable}. CASPER hardware is also frequently found powering the channelization stage of a correlator, with cross-multiplication powered by commodity GPU hardware. This heterogeneous architecture has proven very successful with large-$N$ arrays, where the high arithmetic density allows leveraging of the computational power of GPUs \citep{scalable-gpu-fpga, kocz2015, chime-correlator}. Moreover, the sharing of code, such as the popular GPU-accelerated cross-correlation engine, \emph{xGPU} \citep{xgpu}, has enabled multiple observatories to leverage GPUs effectively with minimal engineering effort.

An example of an ultra-wideband CASPER digital backend using the canonical packetized correlator architecture alongside a phased-array VLBI recording system is SWARM, the SMA Wideband Astronomical ROACH2 Machine, recently commissioned at the Submillimeter Array (SMA). \citep{swarm}. SWARM has recently been deployed as the primary facility back-end at the SMA, and integrates two instruments: a correlator with 140~kHz spectral resolution across its full 32~GHz band, used for connected interferometric observations; and a 64~Gb/s phased array VLBI recording system. SWARM is built with ROACH2 boards with highly-utilized FPGAs running at 286 MHz. In addition to ROACH2 SWARM uses ADC1x5000 digitizers to sample a 2.3 GHz Nyquist band.

SWARM represents the widest bandwidth CASPER correlator currently deployed. In replacing the SMA's previous ASIC-based correlator, it has reduced power consumption by an order of magnitude. The work of the SWARM team, particularly in characterizing the ADC1x5000 \citep{adc5g} and integrating its interface into the CASPER toolflow, has been leveraged by several projects, including the STARBURST, VEGAS and AMI digital back-ends (see Tables \ref{table:casper-instruments-correlators} and \ref{table:casper-instruments-spectrometers}).

\begin{longtable}{ccp{10cm}}
  \caption{Correlators and beamformers using CASPER hardware for either their `F', `X' or beamforming stages.}\\
  Instrument & Year & Description \\
  \hline \endfirsthead
  \caption[]{continued}\\
  Instrument & Year & Description \\
  \hline \endhead
  KAT7             & 2010 & 7 dual-pol antenna full-stokes FX correlator, based on 16 ROACH boards \citep{Foley01082016, man14}. \\
  PAPER            & 2010 & 100~MHz FX correlator originally based on iBOBs, and later upgraded to ROACH, and then ROACH2 boards. Currently supports 256 inputs, using 8 ROACH2 boards for channelization followed by a GPU-based `X' stage powered by LEDA's \emph{xGPU} correlation code \citep{2010AJ....139.1468P, 2014ApJ...788..106P, paper64a}.\\
  ATA              & 2011 & 42 dual-pol antenna beamformer for SETI searches, capable of forming 3 beams with 100~MHz bandwidth. Implemented using 48 iBOBs with iADC digitizers and 15 BEE2 boards\citep{ata-beamformer}.\\
  LEDA             & 2012 & 58~MHz, 512-input digitization, channelization and packetization system for a GPU correlator backend. Implemented using 16 ROACH2 boards with 32 ADC16x250-8 digitizers \citep{doi:10.1142/S2251171715500038}.\\
  ARI              & 2012 & 21-cm dual-antenna interferometer for teaching purposes. Based on a single ROACH and ADC2x1000-8 \citep{MScSalas2014}. \\
  MAD              & 2013 & 16~MHz, 18-input FX correlator and beamforming system for low-frequency array prototyping for the SKA. Implemented using a single ROACH board with ADC64x64-12 digitizer \citep{Pupillo2015, RDS:RDS20336}. \\
  pocketcorr       & 2014 &  Multi-platform (ROACH, ROACH2, SNAP) single-board FX correlator. Used in HYPERION deployment and PAPER testing\footnote{\url{https://github.com/domagalski/pocketcorr}} \\
  Medicina FFTT    & 2014 & A digitization, channelization, beamforming and correlation system, used to demonstrate direct-imaging on the BEST-2 Array. Based on 3 ROACH1 boards and an ADC64x64-12 digitizer \citep{Foster11042014}. \\
  GMRT             & 2014 & 400~MHz bandwidth 32 input correlator for the Giant Metrewave Radio Telescope (GMRT). Various implementation exist, with the latest utilizing 8 ROACH boards with 16 iADC digitizers for digitization, channelization and RFI filtering. Cross-correlation is implemented using GPU hardware \citep{gmrt}\footnote{\url{http://www.ncra.tifr.res.in:8081/~secr-ops/sch/c31webfiles/GMRT_status_doc_June2016.pdf}}. \\
  MITEoR           & 2014 & 50~MHz bandwidth, 64 dual-polarization antenna FX correlator, used to investigate spatial-FFT correlation methods. Implemented using 4 ROACH2 boards each with ADC64x64-12 digitizers \citep{2014MNRAS.445.1084Z}. \\
  AMI              & 2015 & 5~GHz, 4096 channel FX Correlator for the 10-antenna and 8-antenna Arcminute Microkelvin Imager arrays. Based on 18 ROACH2 boards with 36 ADC1x5000-8 digitizers \citep{Zwart21122008, ami-digital}. \\
  MeerKAT AR-1     & 2016 & MeetKAT Array Release 1. Beamformer and correlator system operating between 900 and 1670 MHz with a digital bandwidth of 856 MHz\footnote{\url{http://public.ska.ac.za/meerkat/meerkat-schedule}} \\
  FLAG             & 2016 & 19 dual-polarized input PAF system for the GBT. Provides 150 MHz BW using 5 ROACH2 boards \citep{gb_flag, gb_beamformer}. \\
  BIRALES          & 2016 & 14~MHz bandwidth, 32-input digitization and channelization system for software beamformer. Implemented using a single ROACH board with ADC64x64-12 digitizer \citep{7180719}. \\
  Starburst        & 2016 & 5 GHz, single-baseline FX correlator, based on 4 ROACH2 boards with 8 ADC1x5000-8 digitizers\footnote{See \url{http://www.tauceti.caltech.edu/starburst/Home.html}}. \\
  AMiBA            & 2016 & Upgrade of the AmiBA wideband analog correlator. 7 dual-pol antenna 4.48~GHz bandwidth FX correlator, implemented on 7 ROACH2 boards with 14 ADC1x5000-8 digitizers \citep{amiba-adc, amiba-interim}. \\
  EOVSA            & 2016 & Kurtosis spectrometer and FX correlator for the dual-polarization, 16-antenna Expanded Owens Valley Solar Array. Built using 8 ROACH2 boards each equipped with a pair of katADC digitizers \citep{eovsa}.\\
  SWARM            & 2016 & 32~GHz, 16-input correlator and beamformer for the Submillimeter Array. Implemented using 34 ROACH2 boards and 64 ADC1x5000-8 digitizers \citep{swarm}. \\
  MeerKAT          & Development & ``Facility Instrument'' capable of producing various data products over 856~MHz bandwidth. Modes include 32k channel, 64 dual-pol antenna correlator, beamformer and transient buffer. Currently based on ROACH2 boards, with a SKARAB upgrade forthcoming \citep{meerkat-req}. \\
  HERA             & Development & 100~MHz bandwidth, 700-input FX correlator, constructed using $O(100)$ SNAP boards for digitization and channelization. `X' stage will be carried out either on GPU- or FPGA-based platforms, depending on availability and cost \citep{2016arXiv160607473D}. \\
  \label{table:casper-instruments-correlators}

\end{longtable}

\begin{figure}[ht]
 \centering
 \includegraphics[width=0.3\textwidth]{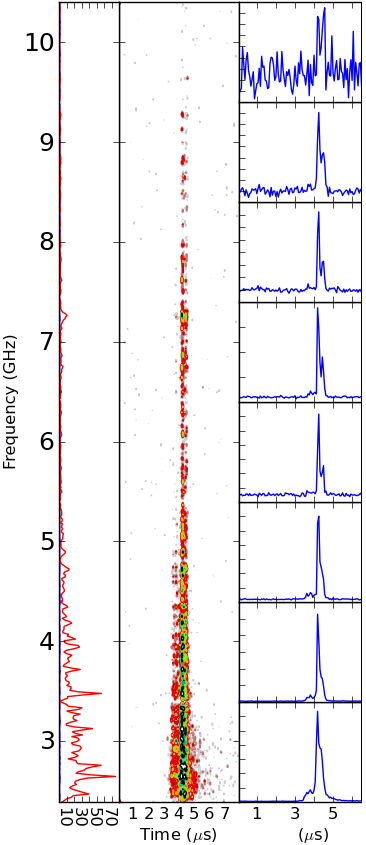}
 \caption{A giant pulse fom the Crab pulsar, captured with the GAVRT wideband recording system. The left panel shows the on-pulse spectrum, the center shows the spectrogram, and the right panels show the integrated pulse in each 1 GHz band.}
 \label{fig:crab-wide}
\end{figure}

The wide range of instruments based on the work of the CASPER collaboration has yielded numerous scientific results. The high-speed R2DBE data recorder \citep{r2dbe} and beamforming systems of the CARMA and SMA arrays are a key part of the Event Horizon Telescope \citep{eht1, eht2}. The GUPPI system deployed at the Robert C. Byrd Green Bank telescope was responsible for discovery of the two-solar-mass neutron star, J1614-2230 \citep{neutron-star}, and continues to power pulsar observations at the GBT including the nanoGRAV pulsar timing project \citep{nanograv}. The Deep Space Network's Goldstone Apple Valley Radio Telescope (GAVRT) has enabled impressive wideband observations using a CASPER backend \citep[Figure~\ref{fig:crab-wide}]{crab-wide}. The BPSR system deployed on the Parkes telescope powered the High Time Resolution Universe Pulsar Survey that discovered a number of new pulsars and other radio transients \citep[for example]{keith2010high, bates2011high, pop-frbs}, and has since been updated with the HIPSR instrument, also powered by CASPER hardware \citep{hipsr}.
The packetized correlator design pioneered by the CASPER group \citep{parsons2008scalable} has powered multiple generations of the Precision Array for Probing the Epoch of Reionization (PAPER) and continues to place field-leading constraints on reionization \citep{paper64a, paper64b}.

\section{Future Directions \& Challenges} \label{sec:Future}
The scale of adoption of CASPER hardware, tools, and architectures has been a great success. However, as a new generation of hardware is released and will likely be used by numerous observatories, it is prudent to cast a critical eye on the work of the collaboration over the past decade. Some legitimate criticisms are:
\begin{enumerate}
    \item The collaboration relies heavily on MATLAB. The closed-source nature of MATLAB is both opposed to the philosophy of the collaboration and places a critical part of the CASPER toolflow outside the control of developers. Further, the dependence of the toolflow on MATLAB places a significant licensing cost on users.

    \item There is healthy reuse of DSP libraries by users of the CASPER toolflow. However, these are not easily utilized by those wishing to build instruments using standard FPGA vendor tools, owing to their strong coupling to the MATLAB Simulink environment.

    \item With so many active developers, maintaining core collaborator libraries that are simultaneously bug-free and incorporate features implemented by different institutions has proven challenging. This issue is exacerbated by the binary-like nature of CASPER's core Simulink libraries, which are not easily managed by version control tools.

    \item The Simulink environment utilized by CASPER is sensitive to software version changes. This brittleness directly impacts the ability of collaborators working at different institutions to easily share designs and contribute code to core CASPER repositories.

    \item Historically, only very few developers possess the necessary knowledge to add features to the CASPER toolflow and integrate new hardware.
\end{enumerate}

The last of these issues should be alleviated with adoption of the JASPER flow. Though still in its infancy, its Python implementation is designed to have a much lower barrier to entry than the original MATLAB implementation. The new toolflow also aims to solve other issues associated with the MATLAB Simulink environment, by providing a route to utilizing other design entry front-ends.

Unfortunately, much of the DSP library development of the last decade is inextricably tied into the Simulink design tool. Creating a set of libraries which are truly agnostic of the design environment is a key step in increasing the flexibility of the CASPER ecosystem. Such a library need not necessarily be developed from scratch, and CASPER is keen to leverage the wealth of DSP modules which exist both inside and outside the radio-astronomy community.

Most importantly, the collaboration must be aware of the changing landscape of digital computation. Today instruments can be built with FPGAs which in the past would have demanded ASICs. It is already clear that in many scenarios GPUs are capable of filling roles where CASPER users have previously required FPGAs. With this in mind, CASPER must apply its principles of modularity, flexiblity and reusablility to the ever-growing collection of software resources developed by the community.

\section{Conclusions} \label{sec:Conclusions}
Over the past decade over 500 CASPER FPGA-boards have been delivered to collaborators who have used them to build more than 45 instruments worldwide. These instruments serve extraordinarily wide-ranging purposes - from single-board educational tools to peta-op-scale correlators to multi-functional facility instruments.

The newest generations of CASPER hardware -- the SKARAB and SNAP platforms -- have catalyzed the development of a new CASPER toolflow, distinct from the bee\_xps flow on which the collaboration has relied for over a decade. This new flow, JASPER, aims to lay the foundations for flexible support of multiple FPGA design tools, alleviating some of the limitations of the current MATLAB/Simulink environment while maintaining backwards compatibility with current designs.

The flexible DSP libraries which have been vital in enabling the design re-use and modularity championed by the collaboration are not compatible with a CASPER ecosystem which does not include Simulink. Designing (or co-opting) a new open-source DSP library on which to base future designs remains a key hurdle for the future.

The collaboration continues in its goal of reducing the cost of building radio-astronomy instruments, and in future looks to increase its efforts in developing and maintaining flexible software (CPU and GPU) resources to compliment existing FPGA infrastructure.

\section{Acknowledgments}
CASPER has been supported by National Science Foundation grants 0243040,  0619596,  0906040,  1006509,  1106045, and 1407804. The collaboration gratefully acknowledges the donations of FPGA chips and design tools by Xilinx, via the Xilinx University Program.

\bibliographystyle{ws-jai}

\bibliography{casper-2016}

\end{document}